# Reliability is a new science —— we are on the right way

Xiao-Yang Li [a], Shi-Shun Chen [a], Waichon Lio [a], Rui Kang [a,b,*]

[a] School of Reliability and Systems Engineering, Beihang University, Beijing, China
[b] Smart Aviation Center, Hangzhou International Innovation Institute of Beihang University, Beihang University, Hangzhou, Zhejiang 311115, China

Email:

leexy@buaa.edu.cn (Xiao-Yang Li)

css1107@buaa.edu.cn (Shi-Shun Chen)

liaowj18@buaa.edu.cn (Waichon Lio)

kangrui@buaa.edu.cn (Rui Kang)

Corresponding author[*]: Rui Kang

**Abstract**

Reliability has long been treated as an engineering practice supported by testing, statistics and standards, yet its status as a scientific discipline remains unsettled. From a philosophical perspective, scientific truth is characterized by a dual-structure that links empirical truth and mathematical truth, which requires an axiomatic system that is symbolically expressible and verifiable by universally repeatable controlled experiments. Building on this criterion, this paper examines whether reliability satisfies the dual-structure of scientific truth. Firstly, we analyze the philosophical foundations of the reliability problem, tracing its transition from experiential confidence and engineering practice toward scientific inquiry. Then, reliability science principles are introduced as an axiomatic system consisting of margin, degradation and uncertainty, which define reliability as the repeatability of system performance across time and space. Next, we present reliability science experiments as the empirical aspect of the dual-structure, where controlled and repeatable interventions are designed to verify the causal relations implied by the axioms. Furthermore, we develop the mathematical framework of reliability as the symbolic aspect of the dual-structure, articulating reliability laws through distance, relation and change, and developing a time-dependent measure, Biandong Statistics, to represent varying uncertainty beyond static descriptions. Accordingly, we argue that reliability is indeed a scientific discipline. The applicability of reliability science is demonstrated across engineering, living and social systems, and a unified logic for guiding engineering activities across the entire product lifecycle is provided, linking reliability to the conceptual, development, procurement, production and operation phases within a model-based structure. Finally, a disciplinary big picture of the reliability science is given. Overall, this work establishes reliability science as a coherent scientific discipline whose axioms, experiments and mathematical formalism jointly realize the dual-structure of scientific truth.

# 1 Introduction

The concept of reliability in engineering emerged gradually as manufacturing methods evolved from craftsmanship to mechanized mass production. In early industrial practice, product performance was ensured mainly by overdesign and meticulous workmanship. While effective for small-scale production, this approach could not meet the requirements of large-scale, standardized manufacturing. The introduction of interchangeable parts in the late nineteenth and early twentieth centuries marked a turning point [1-3]. Interchangeability required consistent manufacturing accuracy, which in turn led to the development of statistical quality control. Sampling inspection and probability-based analysis became essential tools to manage process variability. A notable example came from firearm manufacturing, where the need for precise interchangeability promoted the creation of tolerance systems and formal inspection procedures. These methods were effective in reducing variability at the point of production, but they focused exclusively on conformity to specification rather than long-term functional performance.

During the Second World War, quality control techniques were widely applied to military production [4]. However, operational experience revealed a different class of problems. Even components produced under strict quality control exhibited wide variation in failure times during service. In complex military equipment, such as airborne electronics, some units failed after short use while others operated much longer under the same conditions. This inconsistency indicated that product quality at delivery was not sufficient to ensure dependable performance throughout the service life. The challenge shifted from preventing defects at production to ensuring sustained operational performance, which required new concepts and analytical methods.

At this stage, probabilistic models were incorporated into reliability analysis. Lusser introduced a formula expressing system reliability as the product of the reliabilities of its independent components [5]. Originally developed for missile hit probability, the Lusser product formula demonstrated that system reliability decreases rapidly as the number of components increases, unless each component's reliability is exceptionally high. This insight provided a quantitative link between component-level performance and system-level outcomes, and it highlighted the limitations of quality control when applied to complex systems.

Recognizing the strategic importance of this issue, the United States Department of Defense formed the Advisory Group on the Reliability of Electronic Equipment (AGREE) in 1957. The AGREE report defined reliability as the probability that a system performs its intended function without failure for a specified period under stated conditions, clearly distinguishing it from product quality [6]. More importantly, it established that reliability must be addressed during the design phase rather than as a post-production concern. The report also presented a complete framework for reliability engineering, including the definition of reliability metrics, the allocation of reliability requirements across subsystems, the use of prediction methods, the implementation of structured testing and evaluation procedures, the assessment of economic trade-offs, and the inclusion of reliability-related clauses in procurement contracts [4, 7]. The AGREE framework became the foundation for a series of detailed technical standards, such as MIL-STD-781 for reliability testing [8], MIL-STD-217 for reliability prediction [9], and MIL-STD-338 for maintainability analysis [10]. Over time, these standards were transformed into the MIL-HDBK series of handbooks, reflecting a shift from mandatory compliance to authoritative reference. This historical progression, which began with the pursuit of manufacturing consistency, advanced through the probabilistic treatment of system performance, and culminated in a structured and comprehensive approach to reliability, marked the formal establishment of reliability engineering as an independent and systematic discipline.

The codification of reliability engineering through the AGREE report and subsequent standards built a comprehensive set of methods for reliability management in the design stage. However, their scope remained essentially engineering oriented. They offered structured procedures and empirical tools but did not constitute a unified scientific theory of reliability. In practice, their predictive accuracy depended heavily on historical data, idealized statistical models and simplifying assumptions about failure processes. As systems became more complex, operated under more varied conditions and incorporated new technologies such as advanced electronics and software, these assumptions often proved inadequate. The limitations of the engineering framework naturally prompted efforts to explore whether reliability could be grounded in a formal scientific discipline with universal principles and experimentally verifiable laws.

The search for a scientific foundation is often traced to Gnedenko [11], one of the earliest scholars to treat reliability as a subject of theoretical science rather than purely engineering practice. In his formulation, a reliability problem is defined within a formal "reliability space" in which three elements, namely the system states, the time domain and a probability measure, are jointly specified. This construction parallels the axiomatization of probability theory by Kolmogorov and provides a logically self-consistent basis for describing system behavior under uncertainty. By abstracting away from specific hardware, Gnedenko's framework made it possible to treat different systems within a unified mathematical structure. However, his approach remained fundamentally statistical. It characterized reliability in terms of probabilistic distributions of time to failure, without explicitly linking these probabilities to the underlying physical mechanisms.

Later studies attempted to establish reliability measures on the basis of material and functional degradation mechanisms, generally referred to as reliability physics or the physics of failure. Feinberg [12] represents a prominent example of this trend. Although the term "reliability science" was not explicitly stated, he interpreted failure mechanisms and accelerated testing within the framework of irreversible thermodynamics. By modeling degradation mechanisms such as fatigue, wear, oxidation, and chemical reactions through entropy generation, his approach connected measurable physical quantities to service life and provided a theoretical basis for accelerated testing under elevated stress. This brought physical interpretability to reliability indicators, yet the analysis ultimately returned to probability metrics, such as the hazard-rate and the bathtub curve, and thus remained essentially statistical in nature. The same orientation is evident in other reliability physics methods. In 1961, engineers at Bell Laboratories found that the Arrhenius equation, originally formulated to describe the dependence of chemical reaction rates on temperature, could be applied to evaluate the lifetime of semiconductor devices [13]. Since the model shows that higher temperatures accelerate reaction rates and thus shorten device life, it provided a clear rationale for conducting accelerated life tests and opened the way for the systematic use of acceleration models in reliability engineering. Other examples include the Eyring model [14], which incorporates multiple stress factors such as temperature and humidity, and the Coffin-Manson relation for thermal fatigue [15]. Although these models are grounded in physical reasoning, in engineering applications they are used primarily to extrapolate time-to-failure distributions from high-stress laboratory conditions to expected field performance. Their focus is therefore on estimating service life under given conditions rather than on identifying reliability laws that are universal in scope and verifiable through controlled and repeatable experiments.

Rocchi [16] followed another approach by introducing concepts from statistical mechanics into reliability theory. He proposed a Boltzmann-like entropy formalism to describe the evolution of system degradation over time, with particular attention to explaining the empirical bathtub curve of hazard rate behavior. In his interpretation, the three phases of the bathtub curve, which are infant mortality, constant failure rate and wear-out, can be viewed as analogous to thermodynamic phases governed by entropy changes. This analogy offered a narrative link between physical degradation and statistical hazard functions. However, the bathtub curve itself cannot be regarded as a universal scientific law because its shape is inferred from aggregated field data, is sensitive to the mix of underlying failure modes, and cannot be reproduced in a consistent manner under controlled and repeatable laboratory conditions across different systems.

As early scientific explorations of reliability accumulated yet remained fragmented and limited in scope, the academic community began to ask whether reliability could be defined not only as an engineering practice but as a scientific discipline with its own principles and laws. This question came into sharp focus at the 10th International Conference on Mathematical Methods in Reliability in Grenoble in 2017. The organizers convened a dedicated panel session titled "Is Reliability a New Science?", chaired by Singpurwalla. His lead article in the subsequent special issue of Applied Stochastic Models in Business and Industry provided the intellectual anchor for the discussion and synthesized its key themes [17]. In this article, he traced the question back to Gnedenko's foundational contributions [11], framed the definitional and methodological challenges facing the field, and set out the philosophical and practical stakes of recognizing reliability as a science. The special issue compiled six further contributions addressing definitions, philosophy, and methodology [18-23]. Among these contributions, Anderson-Cook [18] examined definitions and concluded that reliability can meet common criteria for science while also sharing features with engineering. Lawless [19] placed reliability closer to engineering but emphasized the central role of statistics for learning and action. Natvig [20] argued that the label matters less than the field's societal importance. Rykov [21] revisited Gnedenko's view and saw reliability as a path to hard knowledge rather than a mature science]. In a rejoinder, Singpurwalla [22] synthesized these positions and emphasized reliability's role as a framework for decision-making under uncertainty, drawing on Bayesian inference as a unifying principle. He also highlighted the philosophical analysis by Zhang et al. [23], which proposed that science embodies a unity of truth and value, where "truth" refers to the verifiable fact of failure and "value" to the imperative of prediction and prevention. Besides, Zhang et al. [23] further proposed that reliability possesses its own historical, theoretical and practical discourses. Historically, it has evolved from implicit concern with failures to an explicit paradigm addressing their mechanisms and prevention. Theoretically, they formalized reliability through three core equations describing performance, performance margin and a reliability metric, thereby linking loss of function to underlying causes and quantifying reliability. Practically, they described reliability as a complete methodological system for coping with failures and their uncertainties, one that has progressed from experience-based avoidance, to semi-quantitative methods, and ultimately toward an envisioned phase of fully quantitative control.

The above debate on whether reliability is a science has focused mainly on macroscopic systems grounded in classical physics and engineering logic. At the microscopic scale, where quantum effects prevail, degradation mechanisms and reliability measures can differ fundamentally. Recent work by Sun and his group extends reliability science into this quantum domain. Cui et al. [24] defined a trajectory-based metric through the evolution of probability amplitudes in Hilbert space, which was derived directly from the equations of quantum mechanics. This captures the deviation of a device's actual trajectory from its ideal one, accounting for coherence and interference, and is fully rooted in quantum dynamics rather than macroscopic failure data. Du et al. [25] further elaborated on the conceptual transition from classical to quantum reliability. They discussed how consistent histories, a foundational concept in quantum theory, provided a natural probabilistic framework for describing the operational lifetime of quantum devices. Embedded in quantum formalism, their model establishes quantum reliability as a distinct branch of reliability science, applicable to emerging quantum technologies and governed by microphysical laws. These studies do not replace macroscopic perspectives but complement them, showing that reliability can be defined from first principles across physical scales. This cross-scale definability strengthens the claim that reliability is not merely an applied methodology but a candidate for a universal science.

Overall, these historical developments, theoretical formulations and philosophical debates suggest that reliability has evolved from an engineering methodology into a candidate for a new scientific discipline [26]. This paper builds upon this trajectory to articulate the principles, experiments, mathematical frameworks and applications of what we propose as reliability science.

## 2 The philosophical foundations of reliability science

### 2.1 Philosophical reflection

The problem of reliability did not arise only with the advent of modern engineering. From a longer historical perspective, experiential concerns related to reliability already existed when humans first began to use tools. Whether a tool would be available at a critical moment, whether it could maintain its intended function through repeated use, and whether it would remain stable under different conditions constituted the earliest judgments of reliability at the level of personal truth. At this stage, reliability was not an explicitly named object of inquiry, but rather an experiential confidence embedded in practical activity.

Viewed from the continuity of technological practice, human concern for reliability gradually emerged alongside the use of tools and systems. During the era of handicraft production, however, this concern remained largely confined to the individual level. The stability of a tool's performance primarily affected the efficiency and safety of a single user, and the consequences of instability were limited to the domain of personal experience.

The Industrial Revolution fundamentally altered this structure. With the emergence of mechanized production and mass manufacturing, instability in system performance was no longer an occasional phenomenon within individual experience, but began to appear repeatedly in the form of batches of similar products. At this point, the reliability problem was elevated systematically from personal truth to social truth. Products of the same type exposed similar failure modes at different times and in the hands of different users, so that reliability was no longer a subjective judgment of whether a product worked well, but a concrete problem that required collective recognition and response.

As industrial systems continued to develop, humanity's capacity to transform the world expanded, and the objects involved in reliability extended from individual products to complex systems and engineering infrastructures composed of many subsystems. At this stage, the consequences of system failure were no longer limited to the interruption of local functions, but could trigger cascading effects that influenced broader structures of social operation. Accordingly, reliability as a problem of social truth was continuously amplified, and its importance became increasingly prominent with the growth of system scale and complexity. It was through this historical process that reliability was gradually detached from immediate personal judgment and objectified as a technical problem. This transformation gave rise to a professional practice centered on engineering methods, namely reliability engineering. Through testing, statistical analysis and standardization, engineering approaches provided effective means to cope with large-scale system failures. This stage should not be regarded as a substitute for reliability science, but rather as its necessary historical precursor.

However, as engineering objects and application contexts continued to change, the limitations of relying solely on engineering methods became increasingly evident. Reliability assessment depended heavily on existing failure data, making it difficult to support forward-looking design decisions. In addition, engineering models were often constructed for specific objects and operating scenarios, which constrained their explanatory power and limited their adaptability. More importantly, such approaches failed to address a more fundamental question: why are people still able, and indeed still required, to form stable confidence in system performance under conditions where change and uncertainty are ubiquitous?

It is precisely these internal tensions that drive a further inquiry: can reliability transcend the level of empirical engineering practice and rise to a science with a clearly defined object of study, a coherent system of principles, and a sound methodological foundation? This question marks the transition of the reliability problem from the experiential domains of personal and social truth toward the theoretical domain of scientific truth. How reliability becomes a science is therefore not optional but necessary, and it is the central issue addressed in the following section.

2.2    How a discipline becomes a science

The status of a discipline as a science cannot be judged only by its technical tools or by the usefulness of its results. Building on Jin's philosophy of truth, science may be understood as a bridge that links empirical truth with mathematical truth [28]. Scientific progress is not a simple accumulation of observations but a dynamic interaction among three essential elements: controlled and repeatable experiments, symbolic representation and axiomatic systems [29, 30]. Controlled experiments ensure that claims remain verifiable within the empirical world. Symbolic representation, most often in mathematical form, provides abstraction and logical extension across contexts. Axiomatic systems integrate both into a coherent structure, allowing knowledge to be both demonstrable and generalizable [31]. These features distinguish science from practical techniques, empirical summaries or philosophical speculation.

The history of science demonstrates this interplay in successive stages, as illustrated in Fig. 1. In early human practice, knot-tying represented the first controlled act of recording quantities, and the abstraction of numbers created a symbolic bridge from experience to representation. In ancient Greece, mathematics reached axiomatic form through Euclid's *Elements*. The scientific bridge was gradually shaped when such formal systems were reconnected with empirical observation in astronomy. Kepler's laws of planetary motion, Galileo's telescopic discoveries, and Newton's synthesis of celestial and terrestrial mechanics in the *Principia* exemplify how observation and mathematical formulation were united into a consistent explanatory framework.

However, until the early twentieth century, it remained at the level of *scientific empirical truth* rather than *scientific truth*. Newton's laws of motion and the law of universal gravitation were powerful symbolic expressions whose predictions matched controlled observations with high precision, yet their validity still depended on agreement with observations rather than on universally repeatable experimental foundations. The turning point came with the emergence of relativity and quantum mechanics. Both theories established their axiomatic systems not merely from empirical agreement but from experiments that were controlled, repeatable and universally verifiable. The Michelson-Morley experiment [32] revealed the failure of classical assumptions about space and time; Einstein's relativity reformulated those axioms, and its predictions, such as the deflection of starlight, were confirmed under controlled conditions [33]. Quantum mechanics, developed in the same era, introduced axioms such as the uncertainty principle, which was mathematically expressed in symbolic form [34] and validated through precisely reproducible experiments [35, 36]. These revolutions marked a fundamental transition in the foundation of science, shifting from empirical truth defined by observational agreement to scientific truth grounded in axioms and universally repeatable controlled experiments. Through this transformation, the interplay between empirical and mathematical truth evolved into a higher synthesis.

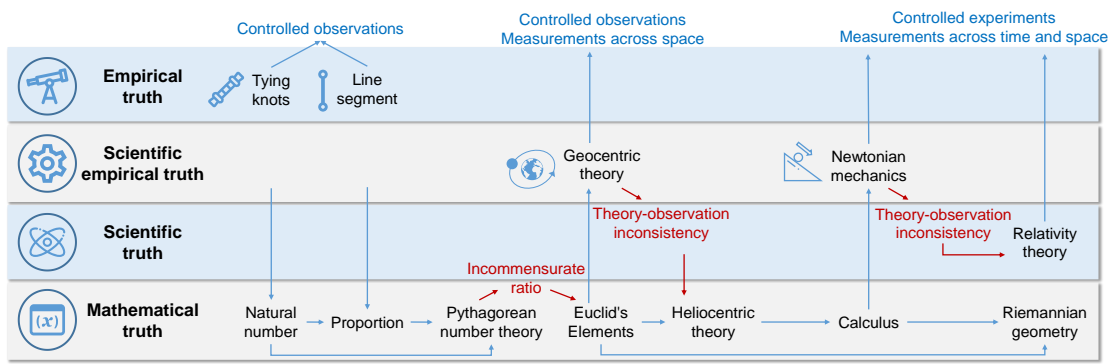

Fig. 1 History of science.

From this perspective, three essential conditions must be met for a discipline to attain the status of science. First, it must possess an axiomatic system, that is, a set of fundamental principles capable of defining its theoretical structure. Second, these axioms must be symbolically expressible in mathematical forms, allowing empirical phenomena to be represented and predicted within a logically coherent framework. Third, the axioms must be verifiable through universally repeatable controlled experiments, ensuring that symbolic representation remains consistent with empirical truth. A discipline that lacks any of these three conditions may still yield valuable practical knowledge or technical applications, but it cannot be regarded as a true science. Science, in its mature form, evolves through the continuous interaction between axiomatic formalization and experimental verification, and it is this dynamic interplay that defines the process of scientific development.

Within this framework, the scientific status of reliability depends on whether it fulfills the above conditions of science. First, reliability must possess a coherent axiomatic system that defines its fundamental principles. Then, the axiomatic system should be dual-structured. On one hand, the axiomatic system must be expressible in symbolic and mathematical form, enabling the laws of reliability to be represented, analyzed and extended beyond specific cases. On the other hand, the axioms must be verifiable through universally repeatable controlled experiments, ensuring that symbolic formulation corresponds to empirical truth. The challenge and promise of reliability science therefore lie in completing this dual-structured axiomatic system, so that reliability can move from being an engineering practice to being recognized as a scientific discipline with both empirical grounding and axiomatic coherence.

## 3 Dual structure of the scientific truth in reliability

### 3.1 Reliability science principle

The first step in demonstrating that reliability qualifies as a science lies in establishing its axiomatic systems, also known as *reliability science principles*. To understand what the reliability science principles truly represent, it is necessary to begin with practice, with the way people experience and expect reliability in real systems. In engineering, reliability embodies the expectation that a system will maintain its intended function under varying conditions and over time. Whether one is operating an aircraft, transmitting data through a satellite link, or using a simple household appliance, the underlying demand is consistency of function when subjected to changing environments, aging and operational disturbances.

From this practical perspective, reliability can be defined as *the repeatability of system function across time and space*. This repeatability is not an abstract assumption but arises from controlled observations in engineering practice, such as monitoring failure times, measuring performance changes and analyzing how systems sustain their function in diverse environments. Reliability science, therefore, must explain not only whether a system performs but also how its performance maintains and why its behavior varies.

Since the repeatability of system function is regarded as the essence of reliability, the first step toward theoretical formulation is to express how the system function can be quantitatively assessed. This is accomplished through the concept of performance margin, which represents the measurable difference between system capability and functional requirement. A system is considered reliable when its performance exceeds the performance threshold of requirements, and unreliable when it falls short. This concept underlies the *Reliability Principle of Margin*, which is the first of the three axioms of reliability science [37]. The corresponding mathematical formal expression is:

$$M = m(P, P_{th}), \qquad (1)$$

where $M$ is the performance margin, $P$ is the measured performance, $P_{th}$ is the performance threshold of requirements determined by user needs or standards, and $m(\cdot)$ is a margin function that quantifies the distance between $P$ and $P_{th}$.

The reliability principle of margin captures the most direct causal law: system reliability depends on the balance between capability and requirement. It is in this sense that reliability connects naturally with management science, where system performance, user expectation and environmental constraint interact dynamically. The equilibrium among these elements determines not only the operational state of a system but also its perceived reliability within a broader socio-technical context.

Practice further shows that the performance margin does not remain fixed. Durability testing and field observations consistently demonstrate that performance degrades as systems are used. This degradation may take the form of wear, fatigue or corrosion, depending on the physical nature of the system. These empirical laws are consistent with the second law of thermodynamics, which states that entropy inevitably increases and that no system can maintain its original functional state indefinitely. The second principle, known as the *Reliability Principle of Degradation*, arises from the recognition that system performance is not static but evolves with irreversible time [37]. The corresponding mathematical formal expression is:

$$P = f_{\tilde{t}}(\mathbf{X}, \mathbf{Y}), \qquad (2)$$

where $\vec{t}$ denotes the irreversible time, **X** represents internal variables such as material properties or geometric parameters, **Y** represents external variables such as temperature, vibration or humidity, and $f_{\vec{t}}$ is the degradation function that integrates these influences.

Based on the reliability principle of degradation, reliability science is closely aligned with the principles of physics, especially thermodynamics and statistical mechanics, which describe the irreversible evolution of systems. At the same time, degradation mechanisms are governed by the specialized laws of different disciplines. Mechanical systems follow the principles of material mechanics and tribology; electronic systems follow the principles of circuit theory and semiconductor physics; control and software systems follow dynamic system theory and algorithmic stability. More generally, from the standpoint of systems science, the same logic applies across different levels of organization, from simple mechanical structures to complex socio-technical systems. This broad correspondence demonstrates that reliability science is not isolated from the established disciplines but integrates their governing principles into a unified explanatory framework.

Even when degradation mechanisms are identified and quantified, repeatability is never purely deterministic. Under nominally identical conditions, failures occur at different times, and measured performances exhibit variation. This leads to the *Reliability Principle of Uncertainty*, which describes the inherent variability of systems and the limitations of human knowledge [37]. The corresponding mathematical formal expression is:

$$R = c\{m_{\vec{t}}(\tilde{P}, \tilde{P}_{th}) > 0\}, \tag{3}$$

where $R$ is the reliability, defined as the possibility that the margin remains positive, $\tilde{P}$ is the performance considered as a random variable, $\tilde{P}_{th}$ is the performance threshold of requirements with uncertainty, $m_{\vec{t}}(\cdot)$ is the margin function accounting for degradation over time, and $c\{\cdot\}$ is a posibility measure such as probability measure or uncertain measure.

In practice, uncertainty arises from two fundamental sources. One is random uncertainty, reflecting natural variability in materials, manufacturing and environments; the other is epistemic uncertainty, reflecting incomplete understanding of mechanisms or insufficient data for inference. Reliability science incorporates both forms by expressing reliability as the possibility that performance remains above the performance threshold of requirements over time. At a deeper level, it also connects the field with fundamental physics and epistemology. Quantum mechanics establishes the intrinsic existence of randomness in nature, while cognitive science and philosophy of knowledge recognize the unavoidable limits of human understanding. Reliability science, therefore, is built upon both the physical foundation of stochastic behavior and the epistemic foundation of reasoning under uncertainty.

3.2    Reliability science experiment

*Reliability science experiment* [38] builds upon the three reliability principles of margin, degradation and uncertainty, and is designed to verify the causal relations they imply. Its purpose is not to gather failure statistics but to examine whether the relations among system performance, performance requirements and margin hold under controlled and repeatable conditions. A reliability experiment begins by identifying the internal variables that influence system performance, such as material properties, structural design or component interaction, together with the external variables represented by temperature, loading and environmental stresses. Then, the irreversible time is incorporated, which governs the evolution of system performance. By controlling and intervening these factors precisely, the experiment aims to reproduce the same pattern of margin change and performance degradation across independent trials, which is the defining feature of scientific repeatability.

The internal coherence of reliability experiments depends on three conceptual foundations: law clarity, black-box epistemology and the causal chance relation. Law clarity requires that an experiment begin with a concrete hypothesis about how the margin degrades, whether in the form of a quantitative model or a qualitative conjecture, and end with a validated or refined reliability model. Black-box epistemology helps determine the appropriate type of experiment based on the alignment between the system's controllability and its modeling level. For complex systems or those with limited controllability, black-box experiments are used to observe input-output behaviors. When the modeling and control levels are consistent, white-box experiments are feasible. In cases of mismatch, equivalent updating is used to transform the experiment into a gray-box type. This adaptation ensures reproducibility of experimental conditions and results, even in small-sample scenarios. The causal chance relation ensures experiments to not only verify the relationship between deterministic margin and performance but also to quantify various uncertainties such as randomness and cognitive factors, thereby ensuring that experimental conclusions better reflect engineering practice.

These conceptual foundations are carried into practice through principles of system integration, classification judgment and optimization equilibrium. The system integration principle aligns with the epistemology of the black box, emphasizing that reliability experiment should be conducted on the system as a whole, so that the observed behavior reflects the combined effect of components and their interactions. For example, when examining the reliability of a mechatronic device, it is not enough to test the degradation of a single part. The flow of energy and information through the system should also be captured in the margin model to avoid misleading conclusions drawn from isolated analysis. The classification judgment principle connects law clarity with the causal chance relation. Deterministic relations are examined by controlling variables at specified levels and observing the resulting changes in performance, while uncertainties are quantified through random sampling or multi-scenario testing that reveals the distribution and influence of stochastic and cognitive factors. The validation logic is further adjusted according to whether the experiment adopts a black-box, white-box or gray-box approach, so that the level of control and modeling remain consistent across experimental types. The optimization equilibrium principle resolves the practical tension between universal repeatability and engineering constraints. In black-box settings, for example, the determination of sample size must consider the improvement in uncertainty quantification that comes from additional samples, while also accounting for the cost and feasibility of conducting large numbers of tests. Similar considerations apply to the choice of stress levels, observation intervals and test duration. By balancing scientific rigor with practical limitations, the experiment remains both credible and implementable.

Through the combination of these principles with the conceptual foundations, reliability science experiments are able to verify causal laws with precision while providing robust support for reliability design, testing and evaluation in engineering practice.

3.3  Reliability mathematical framework

The mathematical framework of reliability is grounded in the three reliability principles of margin, degradation and uncertainty. By representing these principles through the mathematical tools of distance, relation and change, it forms a precise symbolic system for expressing reliability laws. This framework provides a rigorous mathematical foundation needed for both experimental verification and engineering application.

Distance quantifies the deviation between a system's actual state and its reliable state, which is described by Eq. (1). While the term originates from classical geometry, its meaning here is fundamentally different. In Euclidean geometry, distance symbolizes the measured separation between two points in physical space, constrained by non-negativity, symmetry, and the triangle inequality. In the reliability mathematical framework, however, distance extends to a functional space defined by the boundary between reliability and failure. It no longer describes spatial separation but measures how system performance deviates from the performance threshold of requirements. A positive distance indicates sufficient performance, a negative one indicates deficiency, and the absolute value reflects the degree of deviation from reliability. By allowing both positive and negative values, the reliability mathematical framework transforms the geometric notion of distance into a functional measure of performance sufficiency.

Relation is a central concept in the mathematical framework of reliability, characterizing system composition and performance degradation, which is described by Eq. (2). Its definition inherits the abstract meaning of relation in mathematics, yet it gains new significance in the context of reliability because the object of study is a system composed of interacting elements whose behaviors cannot be understood in isolation. Accordingly, the mathematical tools employed in this field naturally focus on representing interdependence, most of which are grounded in functional mathematics. These include analysis (e.g., mathematical analysis, real analysis, complex analysis, and functional analysis), algebra, geometry and category theory. The common goal of these branches is to transform the interactive laws among key system parameters into quantifiable and derivable functional relations. Analytical mathematics quantifies the rate of performance degradation as a function of internal and external variables, thereby providing a precise description of the continuous evolution of system performance. Algebra characterizes the structural organization of interactions through abstract symbolic systems (e.g., groups, rings and fields), which can represent coupling relations among components in a concise and consistent manner. Geometry captures positional and morphological associations among elements in space, using topological concepts to describe the invariance of spatial structures and to delineate the boundaries of the reliability domain more precisely. Taken together, these mathematical branches form a coherent structure in which analytical, algebraic and geometric representations jointly describe the quantitative, structural and spatial dimensions of system relations.

Change represents the culminating dimension of the reliability mathematical framework, describing how system performance evolves across time and space. Its essence lies in the dynamic process through which a system moves from a reliable to an unreliable state under the influence of irreversible time. Since the evolution of a system is governed by both random uncertainty and epistemic uncertainty, deterministic models cannot capture the full complexity of its internal and external dynamics. Consequently, an accurate representation of this process must rely on axiomatic mathematical tools that can address uncertainty described by Eq. (3), most notably probability theory and uncertainty theory. The basic tools for describing performance degradation are stochastic process and uncertain process [39, 40]. In a probabilistic framework, classical theorems such as the law of large numbers and the central limit theorem must be satisfied, ensuring that outcomes converge to their expected values when the number of samples increases. However, when data are scarce or when the population itself changes over time, the empirical mean derived from observation can differ markedly from theoretical expectation. The formalism of probability theory nevertheless enforces convergence toward the expected value, which may drive predictions toward erroneous stability and lead to forecasts that are overly optimistic. Uncertainty theory provides a complementary yet opposite perspective. The uncertain process preserves properties such as distribution conservation and variance invariance, which guarantee stability under limited information. When data become abundant, however, these constraints cause the model to diverge, producing results that are overly conservative and often lacking predictive usefulness. The contrast between probability and uncertainty reveals a deeper structural limitation: both frameworks assume a static logical space in which measures remain invariant with respect to time. Yet in real systems governed by irreversible degradation process, both the underlying data distribution and the epistemic interpretation of uncertainty change over time, making static measures inadequate for describing reliability.

To resolve this foundational gap, an extended axiomatic framework termed *Biandong Statistics* is introduced, within which the measure evolves with irreversible time. The essence of this framework lies in addressing a fundamental deficiency of traditional mathematics: by overemphasizing simplicity and closure of rules, classical axiomatic systems have created an unbridgeable gap between the symbolic world and the empirical world. Biandong statistics seeks to fix this gap by defining a new type of mathematical measure situated between probability measure and uncertain measure. Consider a nonempty set $\Gamma$ and a $\sigma$-algebra $\mathcal{L}$ on $\Gamma$. The structure $(\Gamma, \mathcal{L})$ forms a measurable space, where each $\Lambda$ in $\mathcal{L}$ represents an event. A biandong measure Bd is given on $\mathcal{L}$, satisfying the following axioms.

1) Normality axiom: $\text{Bd}\{\Gamma\} = 1$ for the universal set $\Gamma$.
2) Duality axiom: $\text{Bd}\{\Lambda\} + \text{Bd}\{\Lambda^c\} = 1$ for any event $\Lambda$.
3) Subadditivity axiom: For every countable sequence of events $\Lambda_1, \Lambda_2, ...$, we have

$$\text{Bd}\left\{\bigcup_{i=1}^{\infty} \Lambda_i\right\} \leq \sum_{i=1}^{\infty} \text{Bd}\{\Lambda_i\}. \tag{4}$$

To calculate the biandong measures of product events, the following product evolution axiom is proposed. Let $(\Gamma_k, \mathcal{L}_k, \text{Bd}_k)$ be biandong spaces for $k = 1, 2, ...$, the product evolution measure Bd is a biandong measure satisfying

$$\text{Bd}\left\{\prod_{i=1}^{\infty} \text{Bd}\{\Lambda_i\}\right\} = \Phi(\vec{t}) \prod_{i=1}^{\infty} \text{Bd}\{\Lambda_i\} + \Psi(\vec{t}) \bigwedge_{i=1}^{\infty} \text{Bd}\{\Lambda_i\}, \tag{5}$$

where $\Phi(\vec{t})$ and $\Psi(\vec{t})$ are evolving unit functions, which jointly determine how the measure transitions between probability measure and uncertain measure. These functions satisfy $0 \leq \Phi(\vec{t}) \leq 1$, $0 \leq \Psi(\vec{t}) \leq 1$, and $\Phi(\vec{t}) + \Psi(\vec{t}) = 1$ for any irreversible time $\vec{t}$. When $\Psi(\vec{t}) \equiv 0$, the framework naturally reduces to classical probability, thereby recovering countable additivity. This construction transforms the traditional fixed set of axioms into an axiomatic class that evolves with time. The specific functional forms of $\Phi(\vec{t})$ and $\Psi(\vec{t})$ are determined by modeling the evolution process itself, allowing the measure to adapt to empirical conditions and to reflect both stochastic variation and epistemic constraint. Biandong statistics thus mediates between the optimistic tendency of probability theory and the conservative restraint of uncertainty theory. By embedding time dependence directly within the axioms, it provides a unified framework that is both mathematically rigorous and empirically consistent.

### 3.4 Reliability is a science

Jin's philosophy of truth [28] holds that the essence of science lies in forming a closed feedback loop between empirical truth and mathematical truth, with scientific truth serving as the bridge that connects them. The cornerstone of this bridge lies in the universal repeatability of controlled experiments and the symbolic connections of axiomatic systems. Reliability fits precisely with this framework: science principles, scientific experiments and mathematical framework jointly create a rigorous loop of scientific truth, as follows:

The three reliability science principles form an axiomatic system, acting as the logical starting point for mathematical truth. They are not fragmented empirical summaries but abstract extractions of the system's inherent reliability. The reliability principle of margin characterizes the deviation between system performance and its performance threshold of requirements; the reliability principle of degradation reflects the universal law of time irreversibility; and the reliability principle of uncertainty tackles the shared issue of real-world variability and cognitive limitations. These three principles collectively establish a self-consistent axiomatic framework that maintains the alignment of mathematical truth with the essence of reliability, similar to frameworks in the physics.

Reliability science experiment stands as the core embodiment of empirical truth. Through universally repeatable controlled interventions, it establishes a bridge to the axiomatic system. The experiment strictly controls internal variables (such as component materials and structural parameters), external variables (such as environmental stress and working conditions), and the irreversible time. Thus, empirical phenomena such as margin variation and performance degradation can be consistently reproduced under identical conditions. Whether evaluating performance margins and performance degradation of ion thrusters [41], coarse tracking systems [42] or cloud data centers [43], consistent results can be obtained across researchers. Such universal reproducibility characterizes the essence of empirical truth and facilitates the application of abstract axiomatic systems in the real world, providing observable and testable empirical material for mathematical symbols.

The mathematical framework of reliability is the medium that binds axiomatic system to empirical truth using precise symbolic expressions. It transforms the three principles into computable, inferable mathematical formulas. Margin is quantified via distance functions, transforming the empirical judgment of the reliability boundary into a rigorous mathematical metric. Degradation is modeled as a function of time and variables, converting the observed pattern of performance evolution over time into an iterative mathematical trajectory. Uncertainty is quantifies using mathematical measures, allowing both randomness and cognitive limitations in empirical truth to be formalized within a mathematical structure. These symbols are not merely tools for describing empirical phenomena. They also enable predicting unobserved empirical phenomena through logical deduction, such as estimating the remaining useful life of a device based on the degradation function. This process facilitates a dynamic cycle: from axioms to mathematical inference, then to experimental validation, and finally to model refinement. Mathematical symbols help precisely align the axiomatic system with empirical truth, while empirical truth provides the foundation for verifying and refining mathematical predictions. Together, they form the scientific truth of reliability, as shown in Fig. 2.

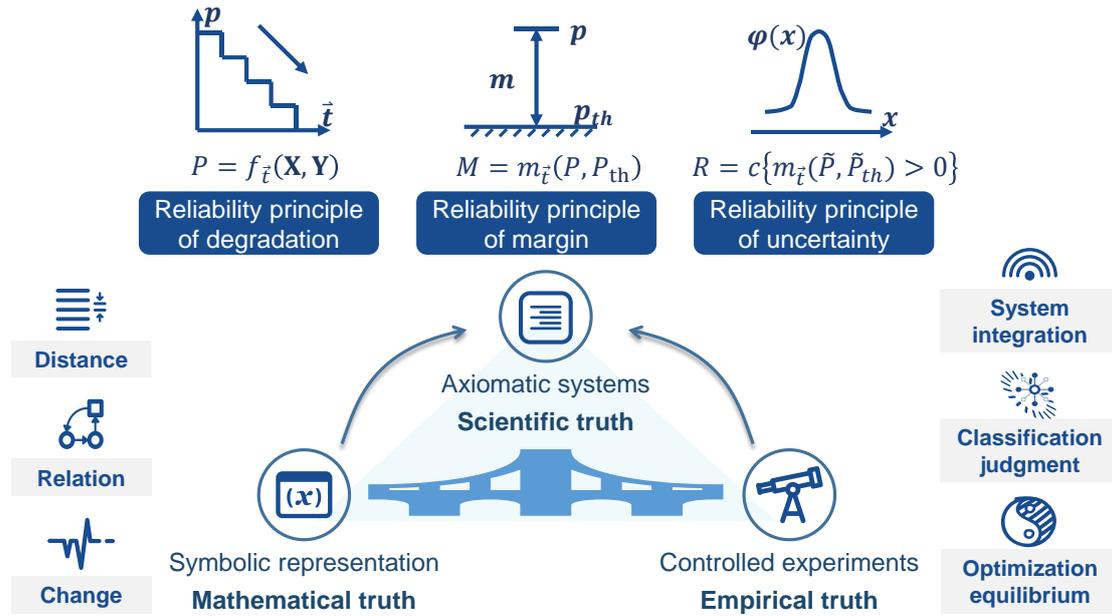

Fig. 2　Dual structure of the scientific truth in reliability

　　In summary, reliability possesses axioms that define its theoretical essence, expresses those axioms through a coherent system of mathematical symbols, and verifies them through universally repeatable controlled experiments. This alignment fully embodies the core criterion that scientific truth is the bridge between empirical truth and mathematical truth. The logical consistency of its theoretical framework, the empirical reproducibility of its experiments, and the precise inferential power of its mathematical expressions collectively demonstrate that reliability is a rigorous science. Belief reliability theory [37] represents one concrete realization of such a reliability science, in which the axiomatic principles, experimental verification, and mathematical formalism are systematically integrated into a unified theoretical framework.

　　Notably, reliability science has a clear correspondence with other sciences, as shown in Fig. 3. It is situated within the domain of the engineering sciences. Its theoretical foundation lies in fundamental sciences such as thermodynamics, statistical mechanics and quantum mechanics, which establish the basic laws governing the behavior of materials, energy and information. Through the synthesis of different engineering sciences within a system, system reliability emerges. From a broader scientific perspective, systems science addresses what the world is composed of, and complexity science explains how the world emerges through interactions. Reliability science, in contrast, focuses on how the world can maintain stable and dependable operation and avoid systemic breakdown. In essence, reliability science arises through the emergent integration of other sciences.

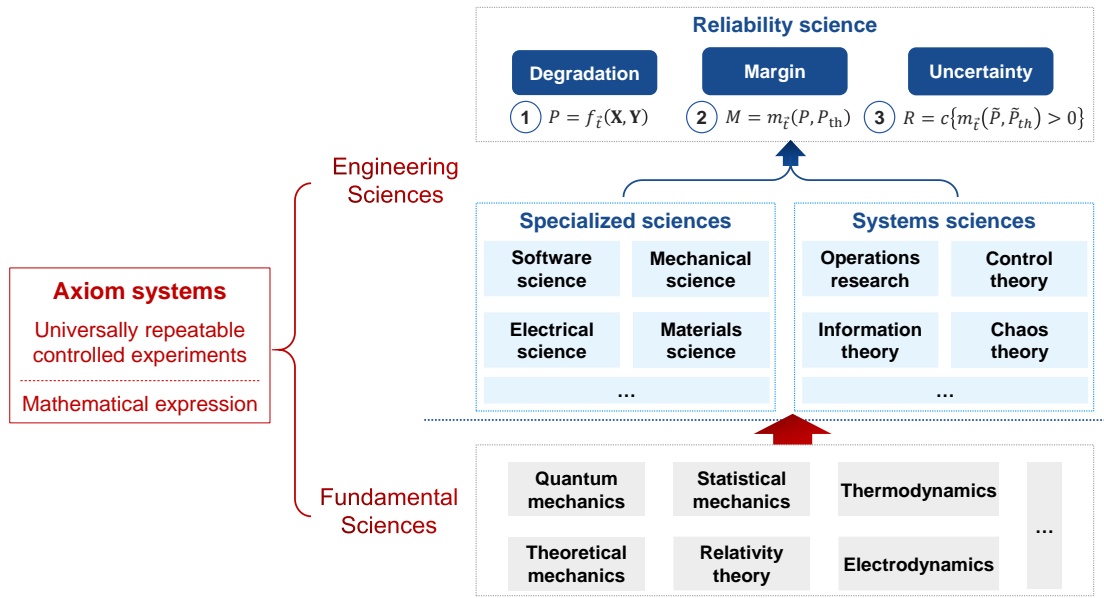

Fig. 3  Relationship between reliability science and other sciences.

## 4  Applications of reliability science

Reliability science demonstrates its maturity not only through its theoretical foundation but also through its wide range of applications. The following survey highlights representative studies using the reliability science principles, covering areas from engineering systems to other complex systems, such as living systems or social systems.

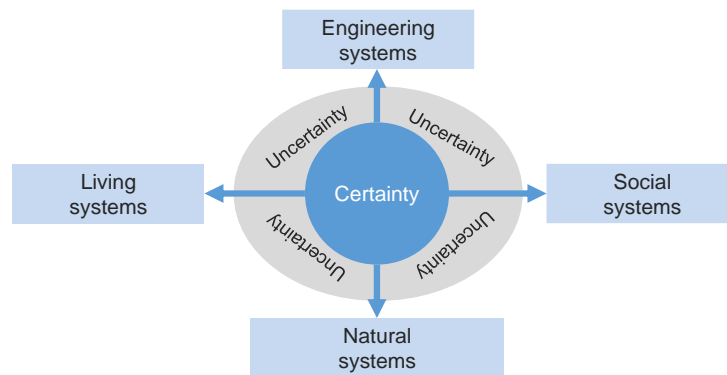

Fig. 4  Application areas of reliability science.

### 4.1  Engineering systems

Engineering systems are human-made constructs developed to fulfill specific functional objectives. In essence, an engineering system represents a purposeful integration of components and relationships that together perform a defined function. As Qian described [44], a system is an organic whole composed of interacting elements that collectively realize a particular function. From this perspective, the foundation of any engineering system lies in its ability to perform functionality.

Mechanical systems provide the most direct link between the principles of reliability science and their functionality. A representative example is the aircraft landing gear lock mechanism [45]. The study defined performance margin based on the locking function and modeled wear degradation through the Archard law. It incorporated four categories of uncertainty: manufacturing deviation, material property variation, environmental stress and threshold uncertainty. Sensitivity analysis identified hinge wear, material hardness and hydraulic stability as the dominant factors affecting reliability. The same logic applies to rotating machinery such as planetary and harmonic gear reducers [46-48]. These studies used hysteresis error as the core performance indicator and considered gear wear as the main degradation mechanism, with manufacturing tolerance, load fluctuation and environmental variation treated as uncertain parameters. The results showed that frictional dissipation and operating history jointly determine the evolution of performance and reliability, offering practical insights for reliability design. When multiple physical processes interact, thermo-mechanical systems provide an expanded view. Research on automotive tubular radiators [49] modeled creep-fatigue coupling under cyclic thermal and pressure loads, defining cumulative damage as the measure of performance degradation. Material parameter variability and limited data were represented using uncertainty theory, forming a framework that combines physics-based modeling with epistemic uncertainty quantification. Similar approaches have been applied in fatigue reliability studies [50], where the performance margin was expressed as the distance between current and critical crack lengths. At a higher structural level, precision assemblies highlight reliability behavior dominated by initial accuracy rather than degradation. In the spaceborne synthetic aperture radar antenna deployment mechanism [51], assembly deviations in hinge alignment and rod positioning reduced surface flatness and pointing precision. Reliability was quantified through performance margins defined by allowable tolerance deviation. Related work on satellite-based phased array antennas [52] further linked initial assembly accuracy with long-term degradation, showing that beam pointing error and antenna gain decline progressively as phase shifters and amplifiers age under temperature stress.

Electrical systems offer another domain where reliability science links physical mechanisms with system-level reliability. A representative study on passive RC filters [53] established the performance margin based on cutoff frequency and signal attenuation, and modeled degradation through resistor oxidation and capacitor electrolyte evaporation. Temperature, humidity and voltage were treated as external variables. The results revealed how measurable shifts in circuit response correspond to the degradation of component properties. Device-level analysis under radiation offers a complementary perspective. For CMOS image sensors [54], reliability depends on how total ionizing dose and single-event effects alter key electrical parameters. The study modeled degradation as the cumulative change of these parameters with radiation exposure. The findings of sensitivity analysis provided suggestions for the design and use processes of the CMOS image sensor for reliability improvement. In complex electronic assemblies [55], reliability science principles have been applied to systems subjected to multiple interacting stressors. The proposed structure-overload-performance method analyzed how mechanical, thermal and electrical factors jointly affect solder joints, optocouplers and MOSFETs. This unified approach provides solutions for system-level reliability assessments when different degradation processes act simultaneously.

As mechanical and electrical subsystems become increasingly integrated, reliability science offers a unified framework for analyzing their coupled behavior. Electromechanical systems connect structural motion with electrical conduction, and their reliability depends on how mechanical contact, geometry and load conditions interact during operation. A representative study on a torsion-spring electrical connector [56] captured how structural geometry, contact pressure and load variation affect the performance margin. Degradation was modeled as the decline of electrical contact quality caused by wear and surface oxidation. This framework supports both reliability assessment and design optimization, ensuring reliable performance throughout the connector's life cycle. Another study proposed an attractor-based method to describe the performance evolution of electromechanical systems [57]. The analysis linked system behavior to generalized coordinates, energy functions and damping properties, showing how external loads and dissipation influence the performance margin. A degradation equation characterized the gradual change of microscopic parameters that lead to macroscopic performance decline. This physically grounded approach provides a general framework for reliability analysis in coupled electromechanical systems.

With the progression of electromechanical systems toward higher autonomy and precision, control systems become essential for maintaining stable operation under varying conditions. Reliability science provides a unified framework linking physical performance, control dynamics and uncertainty representation. A representative study on the coarse-tracking system of a satellite optical communication terminal [42] illustrated this integration. The model took tracking accuracy as the key performance indicator and included the physical properties of sensors, actuators and mechanical parts together with external variables such as radiation, temperature fluctuation, contact stress and vibration. Degradation was modeled through radiation damage in sensors, motor demagnetization and bearing wear, which collectively reduced performance margin. The results showed how control performance evolved under coupled degradation and environmental variation. A related study evaluated the reliability of a proportion integration differentiation feedback control system [58]. It linked the loss of tracking accuracy to time-dependent degradation of sensors and actuators while accounting for multiple uncertainty sources. Further studies extended this framework by focusing on epistemic uncertainty in control dynamics [59].

Building on the stability achieved by control systems, information and software systems extend reliability considerations into the cyber domain. In information systems, reliability extends to encompass the stability of information flow, resource allocation and decision processes. A representative study examined the reliability of cloud data centers modeled as service systems composed of computational resources, network links and scheduling mechanisms [43]. Using a generalized Petri net framework, the analysis defined service reliability through task scheduling efficiency, resource utilization and delay propagation. The results showed how reliability science can be applied to systems that couple physical mechanisms with informational processes. In software systems, reliability depends on how faults are detected, corrected and prevented during development and testing. A representative study proposed a software reliability growth model grounded in reliability science principles [60, 61]. The model incorporated testing coverage as a key factor affecting the performance margin, recognizing that limited coverage allows latent faults to remain and restricts reliability improvement. Performance degradation was represented by the change in failure intensity over time, capturing both the loss and recovery of functional capability during testing. By quantifying epistemic uncertainty through limited statistical data, the framework enabled rational reliability evaluation under incomplete information.

As engineering systems become increasingly coupled across physical domains, reliability science offers a unified framework for integrating multidisciplinary mechanisms (e.g., mechanical, electrical and chemical) within a single analytical structure. Lithium-ion batteries provide a representative example where electrochemical degradation governs long-term reliability. One study modeled capacity fade by combining internal variables such as active lithium content, SEI film density and anode surface area with external variables including temperature, discharge current and depth of discharge [62]. The gradual loss of capacity defined the performance margin and enabled reliability evaluation for life prediction and health management. Research on large-scale battery packs in renewable energy systems further linked cell-level electrochemical behavior with system-level dynamics [63]. Key parameters such as state of charge, open-circuit voltage and internal resistance were coupled with environmental factors like temperature and discharge rate, while the degradation of energy output over time represented system-level performance loss. In extreme environments, multidisciplinary reliability frameworks have been applied to aerospace systems. The thermal protection system of reentry vehicles combines heat conduction, radiation and material ablation with aerodynamic heating [64]. Degradation of thermal resistance under prolonged heat flux was modeled explicitly, forming a coherent analysis of high-temperature reliability. Studies on three-grid ion thrusters showed how aperture wear cause performance degradation in electrical propulsion systems [41, 65]. Reliability modeling coupled geometric and operational parameters such as grid thickness, aperture size, discharge voltage and current, providing a quantitative basis for lifetime prediction and experimental design. At the system scale, reliability science principles have been extended to renewable energy generation. A representative study of wind farms developed a performance margin model based on rotor-equivalent wind speed and analyzed how wind distribution, direction and wake effects influence output power and system reliability [66].

Across these examples, reliability science provides a unifying framework for engineering systems. This unified logic allows reliability to be evaluated across diverse domains, from components and subsystems to cyber-physical infrastructures, showing that the same scientific principles govern reliability throughout the engineering systems.

Beyond its application to different classes of engineering systems, reliability science also provides a unified logic for guiding engineering activities across the entire product lifecycle, as illustrated in Fig. 5. This guidance arises from the reliability disciplinary equations derived from the principles of margin, degradation and uncertainty. These equations establish the causal relationship between reliability and the decisions made at each stage of the life cycle. In the conceptual phase, the performance threshold of requirements are transformed into the required reliability level, which are then distributed to the performance margins of subsystems. In the development phase, the internal variables, such as component materials and dimensions, as well as system relationships, including functional principles and fault-tolerant logic, are designed to meet the allocated margin requirements. In the procurement phase, selections are made for the internal variables that form the system. In the production phase, internal variables and system relationships are realized into physical form. In the operation phase, external variables, the performance threshold of requirements and the irreversible time are specified, assessing the repeatability of system function across time and space. As a result, reliability science links the main tasks of each stage with the resulting system reliability, forming a model-based reliability systems engineering framework. In this way, reliability science offers not only an explanatory theory but also a practical foundation for accurate closed-loop control of system reliability.

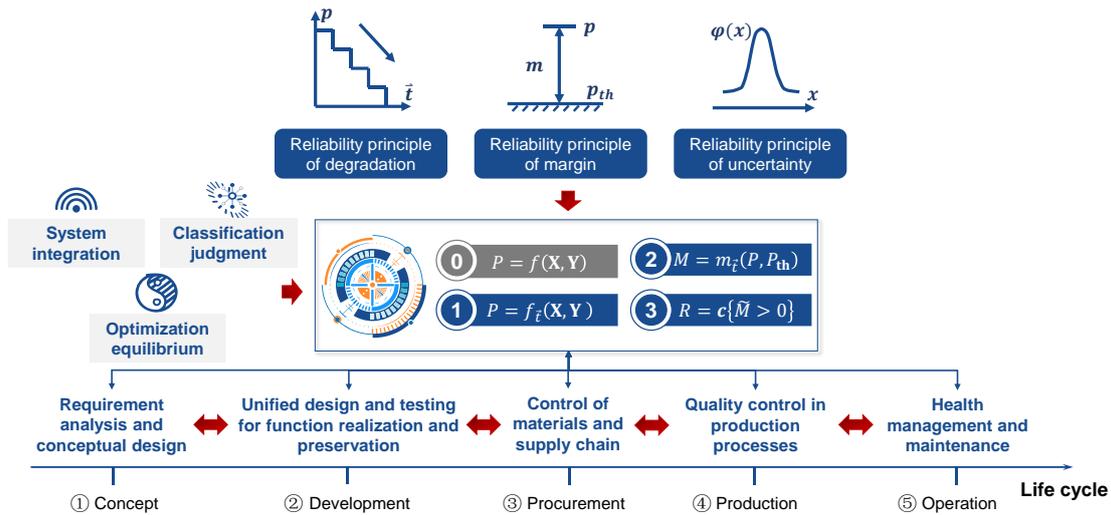

Fig. 5 Guidance of reliability science on the entire product lifecycle.

## 4.2 System extensions

### 4.2.1 Living systems

Living systems differ from engineering systems in that they arise naturally rather than through human design, possessing the intrinsic ability to self-organize and adapt. Their fundamental function is survival, achieved by maintaining internal balance under external and internal disturbances. In human beings, this balance defines health, while its decline manifests as disease and its complete loss results in death. Systems medicine, as developed by Jin and Ling [67], applies systems theory to interpret the human body as a dynamic and self-regulating system. It explains health as a stable state maintained by the coordination of multiple subsystems and views disease as the excessive deviation of this stability. The theory emphasizes that treatment should respect and support the body's inherent capacity for recovery rather than rely solely on external intervention. From this viewpoint, the maintenance of health depends on sustaining sufficient margin to resist perturbation and preserve systemic stability. This understanding converges naturally with the reliability science principles, where performance margin represents the quantitative boundary between normal function, degradation and failure. Thus, systems medicine provides conceptual and empirical evidence that the fundamental logic of reliability, including function, margin, degradation and uncertainty, is equally applicable to living systems.

Following this conceptual connection, recent studies have applied reliability science to specific physiological and pharmacological processes, translating the theoretical principles of systems medicine into quantitative models that describe how biological functions evolve under uncertainty. Liu and Kang established an uncertain differential equation model to describe the pharmacokinetics of digoxin within a two-compartment structure consisting of plasma and tissue distributions [68]. The model incorporated intrinsic physiological processes such as drug transfer and elimination, along with biological noise representing external influences. The decline of drug concentration over time was interpreted as performance degradation, while epistemic uncertainty arising from individual variability and metabolic complexity was quantified through uncertainty theory. Building on this framework, subsequent studies analyzed broader pharmacological processes, including drug absorption and elimination under varying bioavailability and dosage conditions [69], and extended the approach to extravascular administration and nonlinear kinetics based on the Michaelis-Menten model [70]. Across these studies, reliability was defined as the belief degree that concentration remains within the therapeutic range, showing how reliability science unifies pharmacokinetic dynamics with uncertainty quantification in living systems.

Beyond studies on biochemical drug metabolism, another line of research has examined the reliability of physiological regulation at the system level. The human baroreflex provides a representative example, functioning as a closed-loop control process that stabilizes blood pressure through neural and vascular coordination. From the perspective of reliability science, the baroreflex acts as a feedback system that detects deviations in blood pressure, transmits signals through neural pathways, and adjusts cardiac and vascular responses to restore equilibrium. Following this, Shangguan et al. [71] developed a control-based reliability model of the baroreflex, in which regulation error and control effort describe changes in system performance and reserve capacity. The analysis showed that reliability can be represented as the probability that baroreflex regulation remains effective within physiological limits, linking functional stability with control dynamics. In addition to control-based modeling, Li et al. [72] applied entropy methods to evaluate baroreflex regulation under resting conditions, as illustrated in Fig. 6. Using physiological indicators such as baroreflex sensitivity (BRS), heart rate (HR), heart rate variability (HRV) and systolic blood pressure (SBP), they introduced a physiological entropy index to quantify regulatory capacity and systemic uncertainty. Increased entropy in blood pressure indicated weakened stability, whereas reduced entropy in neural or cardiac signals reflected declining function. Together, these studies demonstrate that reliability science provides a coherent framework for understanding how living systems sustain stable performance under uncertain conditions.

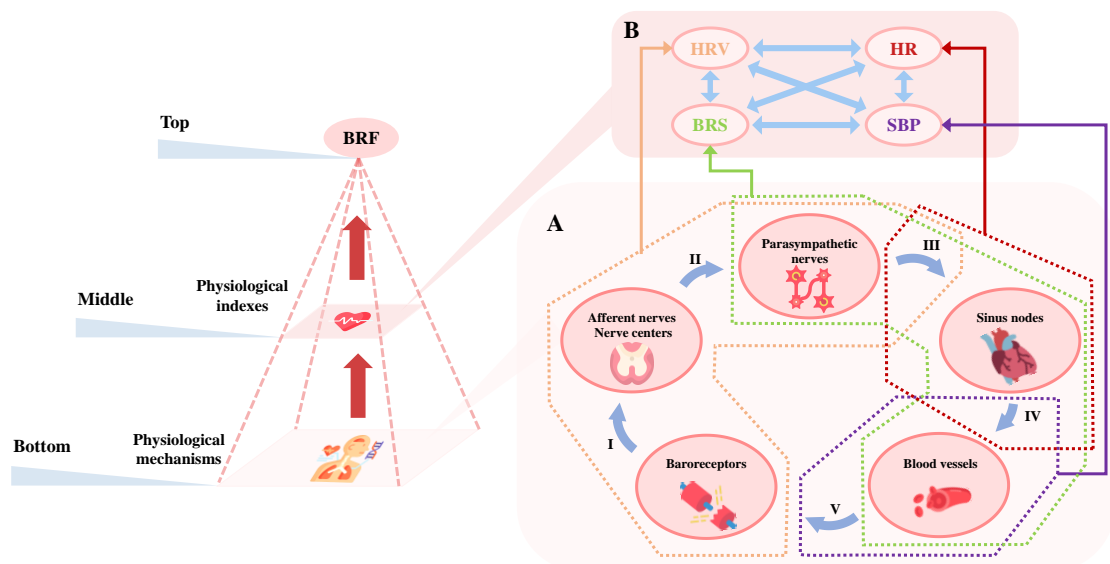

Fig. 6 Emergence from physiological mechanisms to physiological indexes and baroreflex regulation function (BRF) [72]. (A) In the context of physiological mechanisms, nodes represent organs or tissues, while arrows denote physiological processes, where (I) impulses produced by baroreceptors travel via the afferent pathway and are integrated within neural centers; (II) nerve centers activate parasympathetic responses; (III) parasympathetic nerves stimulate the sinus node to change HR; (IV) HR affects BP and blood vessel stretching; (V) baroreceptors sense the stretching. Dashed parts indicate the emergent indexes corresponding to different organs/tissues and processes. (B) Physiological indices, i.e., BRS, SBP, HR, and HRV, emerge from physiological mechanisms. Arrows indicate the possible interactions among indexes.

### 4.2.2 Social systems

In contrast to engineering and living systems, social systems emerge from the collective interactions of individuals, organizations and institutional rules. They operate through communication, cooperation and coordination, enabling resources, information and decisions to circulate within structured networks. The fundamental function of a social system is to sustain organized activity and ensure the continuity of essential functions under changing internal and external conditions. From the perspective of reliability science, the stability of such systems depends on their ability to maintain structural connectivity, functional coordination and adaptive capacity despite disturbances. Performance degradation appears as reduced efficiency, loss of coordination or breakdown of connectivity, while uncertainty originates from incomplete information, behavioral variability and unpredictable external influences. Reliability analysis in this context focuses not on physical failure, but on the persistence of systemic performance under uncertainty, providing a framework to evaluate how social infrastructures such as communication, transportation and demographic systems remain stable and functional in dynamic environments.

Transportation systems provide a representative example, where performance depends on the collective interaction of infrastructure, human behavior and environmental factors. Yang et al. [73] developed an uncertain percolation semi-Markov model that integrated road conditions, traffic demand, driver behavior and signal malfunctions within a unified analytical framework. In this model, performance degradation was described through capacity reduction and signal failure, both influencing system performance margin. By jointly addressing random and epistemic uncertainties, this approach captured the evolutionary behavior of traffic reliability. Subsequent research has examined travel-time reliability as a practical measure of system performance [74]. The results revealed that traffic performance margin was influenced by both regular factors, such as bottlenecks and signal control, and irregular events, including accidents, construction and extreme weather, as illustrated in Fig. 7. Although no performance degradation was involved, fluctuations in travel time effectively reflected the variations of system performance. Beyond transportation, reliability principles have also been extended to population systems [75]. In migration-driven population dynamics, population density was modeled as a state variable affected by external migration sources. The instability or decline of population density corresponded to performance degradation, while uncertainty in migration processes was represented through uncertainty theory. This framework demonstrated how reliability science can be applied to social contexts, where performance was defined by systemic stability rather than physical durability.

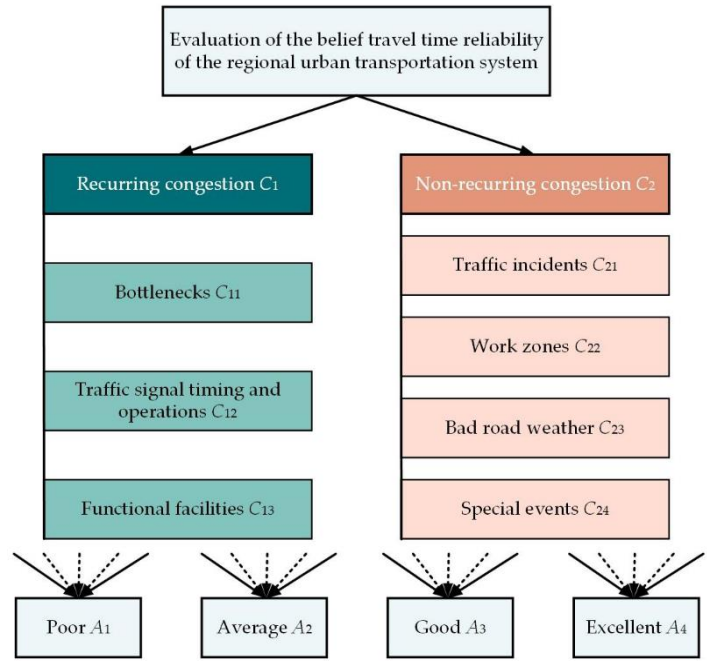

Fig. 7　The hierarchical structure of the belief travel time reliability [74].

## 5　Disciplinary big picture of reliability science

　　Through the preceding analyses across engineering, living and social systems, the universality of reliability science has been empirically demonstrated. Reliability embodies both the empirical authenticity of controlled experiments and the symbolic coherence of an axiomatic system. On this foundation, it becomes possible to articulate the disciplinary big picture of reliability science. Wu [76] has emphasized that every mature discipline should answer a common set of foundational questions : what it studies, the problems it addresses, the ways of thinking it adopts, the analytical methods it employs, and how it relates both to other fields and to the world at large. For reliability science, which is here positioned as an emerging yet independent discipline, such a big picture defines not only its scope but also its intellectual identity. It situates reliability within the broader history of science, clarifies its boundaries, and points toward its future trajectory. Table 1 outlines the disciplinary big picture of reliability science.

Table 1 The Disciplinary Big Picture of Reliability science

| Discipline aspect | Description in reliability science |
|---|---|
| Typical research objects | Everything is a system: the system together with its environment forms a larger system |
| Typical research problems | Systems must be reliable: how systems achieve reliability and how they fail |
| Typical ways of thinking | Systems thinking: integrating internal and external causes, static and dynamic perspectives, capabilities and requirements, certainty and uncertainty |
| Typical analytical methods | Combination of reductionist and emergent approaches: decomposing system failures into element failures, integrating element reliabilities into |
| | system reliability |
| Relationship with other disciplines | Emergent integration of theories from other disciplines |
| Relationship with the world | Establishing a worldview that accepts mortality, a life philosophy that faces uncertainty, and a value system that aspires to lifelong reliability |

The research objects are inherently systemic, recognizing that any system exists within and interacts with a larger system that includes its environment. Accordingly, the central research problem of reliability science is dual in nature: to explain how systems achieve reliable functioning and to understand how and why they fail. The characteristic way of thinking is systems thinking, which unites diverse perspectives: integrating internal and external causes, reconciling static descriptions with dynamic evolution, aligning capabilities with requirements, and balancing certainty with uncertainty. Methodologically, reliability science synthesizes reductionist and emergent approaches. Traditional reliability engineering is predominantly reductionist, analyzing systems by decomposing them into components and tracing failures down to elemental mechanisms at the material or physical level. This decomposition is often carried out through a series of simplifying assumptions that make detailed analysis feasible. However, when results are projected back to the system level, the interactions and dependencies that were simplified or neglected during decomposition are rarely reconstructed explicitly. As a result, system reliability is either treated as a black-box quantity inferred from aggregated component measures, or evaluated without establishing a clear structural link between component behavior and system performance. In this way, while reliability at the system level appears as an emergent outcome, the pathway through which component mechanisms, interactions and uncertainty collectively give rise to that outcome remains unclear. Belief reliability theory provides a theoretical framework in which this emergent process is explicitly represented. It focuses on how system reliability arises from the structured integration of component behavior, interacting mechanisms and multiple sources of uncertainty, rather than assuming system behavior as a black-box outcome. Reliability science, as framed here, synthesizes reductionist analysis with explicit emergence, enabling both detailed component-level investigation and coherent system-level integration. In its relationship to other disciplines, reliability science is the emergent integration of other sciences, which we have discussed in Section 3.4. Its relationship to the world extends beyond technical performance to broader implications. By formalizing how systems sustain function under change and uncertainty, reliability science supports a worldview that acknowledges failure as inevitable, a practical philosophy that confronts uncertainty, and a value orientation that emphasizes dependable operation over the entire lifecycle of systems. In this sense, reliability science is not only a technical discipline but also a scientific framework that connects physical law, system behavior and human purpose. Its disciplinary big picture is therefore a direct reflection of its scientific structure, demonstrating how reliability can be understood, analyzed and designed across domains within a coherent and unified science.

## Declarations of interest

None.

## Acknowledgements

This work was supported by the National Natural Science Foundation of China [grant number

51775020], and the National Natural Science Foundation of China [grant number 62073009].